\documentclass[aps,pra,preprint]{revtex4}
\usepackage{graphics,graphicx,times,bm,bbm,color}

\begin{document}

\title{Reply to M. Ziman's ``Notes on optimality of direct characterization
of quantum dynamics"}
\author{Masoud Mohseni$^{1,2}$ and Daniel A. Lidar$^{2,3}$}
\affiliation{$^{1}$Department of Physics, University of Toronto, 60 St. George St.,
Toronto, Ontario, M5S 1A7, Canada\\
$^{2}$Department of Chemistry, University of Southern California, Los
Angeles, CA 90089\\
$^{3}$Departments of Electrical Engineering-Systems, and Physics, University
of Southern California, Los Angeles, CA 90089}

\begin{abstract}
Recently M. Ziman \cite{MarioNote06} criticized our approach for
quantifying the required physical resources in the theory of
Direct Characterization of Quantum Dynamics (DCQD) \cite%
{MohseniLidar06} in comparison to other quantum process tomography
(QPT) schemes. Here we argue that Ziman's comments regarding
optimality, quantumness, and the novelty of DCQD are inaccurate.
Specifically, we demonstrate that DCQD is optimal with respect to
both the required number of experimental configurations and the
number of possible outcomes over all known QPT schemes in the $%
2^{2n}$ dimensional Hilbert space of $n$ system and $n$ ancilla
qubits. Moreover, we show DCQD is more efficient than all known
QPT schemes in the sense of overall required number of quantum
operations. Furthermore, we argue that DCQD is a new method for
characterizing quantum dynamics and cannot be considered merely as
a subclass of previously known QPT schemes.
\end{abstract}

\maketitle

\section{\protect\bigskip Reply to the first comment: Quantification of
resources}

In general, for a \emph{non-trace preserving} completely positive (CP)
quantum dynamical map acting on a $d$ dimensional quantum system, the number
of independent elements to be characterized is exactly $d^{4}$ \cite%
{Nielsen:book}. In fact, in many important physical situations the quantum
dynamical maps are effectively non-trace preserving. This phenomenon usually
appears either as \emph{loss} for photonic systems (due to inherent
imperfections of optical elements), or \emph{leakage} for atomic and
spin-based quantum systems (due to interactions with photonic and/or
phononic environments, spin-orbital coupling, etc.). Therefore, Ziman's
statement, \textquotedblleft Each quantum device acting on a $d$ dimensional
quantum system is described by $d^{4}-d^{2}$ independent parameters that has
to be specified in arbitrary (complete) process tomography
scheme.\textquotedblright\ \cite{MarioNote06}\ is not completely accurate.
Even though one can always consider a larger Hilbert space such that the map
becomes trace-preserving, this mathematical trick has little or no
physical/practical significance, since in general we do not have full
control over such an extended space. I.e., we cannot arbitrarily redefine
our system and its environment in real life cases.

In Ref. \cite{MohseniLidar06} we only compared our Direct Characterization
of Quantum Dynamics (DCQD) scheme with \emph{separable} quantum process
tomography (QPT) methods, including standard quantum process tomography
(SQPT) \cite{Nielsen:book} and the separable AAPT \cite{DAriano01}. Ziman
raises the possibility of using non-separable QPT methods and proposed new
resource quantification tables (the second and third tables in Ref. \cite%
{MarioNote06}). As we argue below in a detailed analysis of such schemes,
these tables are, unfortunately, incomplete and/or inaccurate.

In general, the required state tomography in AAPT could also be realized by
\emph{non-separable} quantum measurements. These measurements can be
performed either: (1) in the same Hilbert space, such as mutually unbiased
bases (MUB) measurements \cite{Wooters:89,LawrenceMUB02}, or\ (2) in a \emph{%
larger} Hilbert space, such as a generalized measurement or POVM \cite%
{DArianoUQO02}. However, we find that these types of measurements would
hardly have any practical relevance in the context of QPT, because they
require\emph{\ many-body interactions} that are not experimentally
available. In the next two subsections we address each of these two
approaches separately.

\subsection{AAPT with Mutual Unbiased Bases Measurements}

The AAPT scheme utilizes the degrees of freedom of an auxiliary
system, $B$, in order to characterize an unknown quantum dynamical
map acting on a principle system $A$. The information about the
dynamics is obtained by complete quantum state tomography of the
combined system and ancilla. Quantum state tomography in itself is
the task of characterizing an unknown quantum state by measuring
the expectation values of a set of non-commuting observables on
the subensemble of quantum systems prepared in the same state. For
characterizing the density operator of a $d$-dimensional quantum
system, there are in general $d^{2}-1$ non-commuting observables
to be measured. The minimal number of non-commuting measurements,
that corresponds to a mutually unbiased basis, is $d+1$ (for
systems with $d$ being a prime or a power of prime)
\cite{Wooters:89,LawrenceMUB02}.  A set of bases, in a given
Hilbert space, are mutually unbiased if the inner products of each
pair of elements in these bases have the same magnitude.

Let us consider the case of characterizing a non-trace preserving dynamical
map acting on a single qubit $A$, using a single ancilla qubit $B$. For such
a two-qubit system $d=4$, and the number of MUB for the required state
tomography is five. Therefore, the minimum number of ensemble measurements
(experimental configurations) in the AAPT scheme in a single qubit case is
\emph{five}, (as opposed to \emph{four} mentioned in Ref. \cite{MarioNote06}%
). The first measurement provides four independent outcomes and the last
four measurements each provide three independent outcomes. I.e., we have, $%
1\times \lbrack (1\times 4)+(4\times 3)]=16$, where in each term in the sum
the first number ($1,4$) represents the number of required measurements per
input state, and the second number ($4,3$) represents the number of outcomes
for each measurement; $16$ corresponds to the total number of independent
outcomes. This should be compared with $4\times (1\times 4)=16$ \ in the
DCQD scheme \cite{MohseniLidar06}. The first number ($4$)\ is the number of
required different input states (compared to $1$ in the AAPT\ scheme).

It should be noted that if we know the local state of the ancilla (i.e., if
we know the results of the measurements $I^{A}\otimes I^{B}$, $I^{A}\otimes
\sigma _{x}^{B}$, $I^{A}\otimes \sigma _{y}^{B}$ and $I^{A}\otimes \sigma
_{z}^{B}$ from our prior knowledge about the preparation and the fact that
the output state has trace unity), then we need to find only $d^{4}-d^{2}$
parameters for the superoperator. But even in this case, we still need \emph{%
five} different measurement setups. For a simple proof, let us examine the
MUB of a two-qubit system, see the table in Fig.1 (also Ref. \cite%
{LawrenceMUB02}). Obviously, for a non-trace preserving map we need at least
five (ensemble) measurements corresponding to each row (measuring
simultaneously the commuting operators in the first two columns). Now the
question is, how many measurements are needed if we already know the local
state of the ancilla? The answer becomes clear if we note that the
measurements $I^{A}\otimes \sigma _{x}^{B}$, $I^{A}\otimes \sigma _{y}^{B}$,
and $I^{A}\otimes \sigma _{z}^{B}$ appear in the second column of the first
three rows and are redundant. However, we still need to perform all three
(ensemble) measurements corresponding to the first three rows, since the
measurements in the first column (corresponding to the local state of the
principal qubit 1 -- $\sigma _{x}^{A}\otimes I^{B}$, $\sigma _{y}^{A}\otimes
I^{B}$ and $\sigma _{y}^{A}\otimes I^{B}$) do not commute. These three
measurements plus the two measurements related to the 4th and 5th rows
(corresponding to measuring the correlations of the principal qubit and the
ancilla) result in five measurements overall. Note that the above argument
is independent of the basis chosen, since in any other basis the
measurements corresponding to the local state of the ancilla always appear
in different rows due to the non-commuting properties of the Pauli operators.

\begin{figure}[tp]
\includegraphics[width=6.50cm,height=3.8cm]{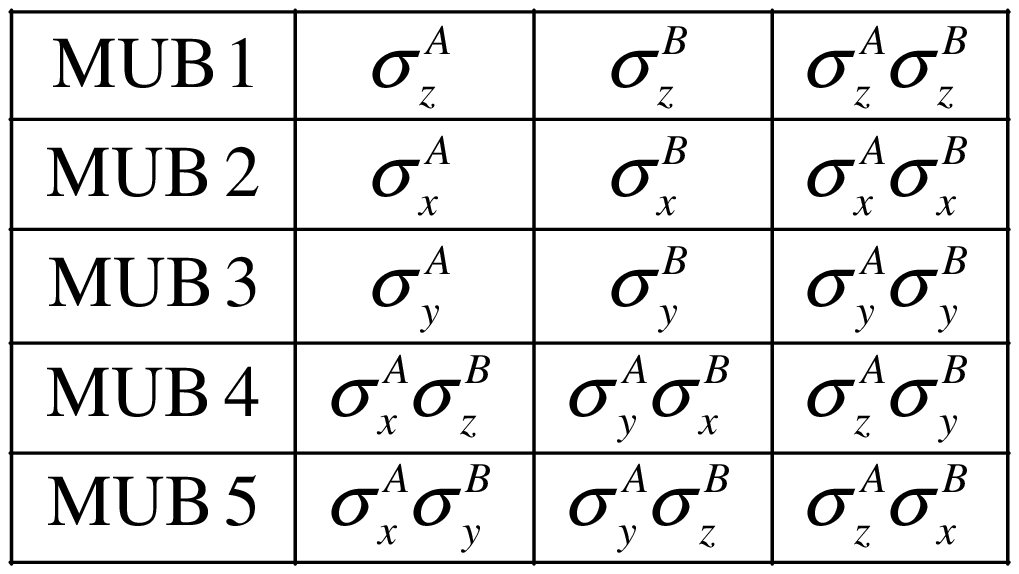}
\caption{A mutually unbiased basis (MUB) of the $4$-dimensional Hilbert
space of the principal qubit $(A)$ and its ancilla $(B)$. Each row
represents a different experimental configuration.}
\end{figure}

For the case of $n$ qubits, and by using $n$ ancillary qubits, the required
measurements for the AAPT scheme can be performed using three different
types of methods: (a) using $16^{n}$ (separable) joint single-qubit
measurements on the $n$ principal and $n$ ancillary qubits, or (b) using $%
5^{n}$ mutually unbiased bases measurements (tensor product of MUB
measurements on two-qubit systems), or more efficiently (c) using $4^{n}+1$
(non-separable) mutually unbiased bases measurements on the Hilbert space of
all $2n$ qubits. The latter method requires \emph{many-body interactions}
between all $2n$ qubits, which are obviously not naturally available. The
required global Hamiltonian can be written as $H=\sum%
\limits_{i=1}^{2n}O_{i}=\sum\limits_{i=1}^{2n}\bigotimes\limits_{k=1}^{2n}%
\sigma _{\alpha (i,k)}^{k}$, where $[O_{i},O_{j}]=0$ for any $i,j\in
\{1,2,\ldots ,2n\}$, such that their common eigenvectors form a MUB, and $%
\sigma _{\alpha (i,k)}^{k}$ is a Pauli operator ($\sigma _{x}^{k}$, $\sigma
_{y}^{k}$ or $\sigma _{z}^{k}$) acting on the $k^{\mathrm{th}}$ qubit. I.e.
one should simultaneously measures $2n$ commuting Hermitian operators $%
O_{i}\in \{O_{1},O_{2},...O_{2n}\}$; each operator $O_{i}$ in itself acts on
$2n$ physical qubits as $O_{i=\{\alpha _{1},\alpha _{2},\alpha
_{3,.}..\}}=\sigma _{\alpha _{1}}^{1}\sigma _{\alpha _{2}}^{2}\sigma
_{\alpha _{3}}^{3}...\sigma _{\alpha _{2n}}^{2n}$ (e.g., for the case of
three qubits see Fig. 2 of Ref. \cite{LawrenceMUB02}).

The operators $O_{i}$ and $O_{j}$ commute globally and are made of tensor
products of Pauli operators, however they cannot be simultaneously measured
locally, i.e. by using only single-qubit devices. The reason is that
according to the Heisenberg uncertainty principle, the outcome of each local
measurement $\sigma _{\alpha }^{k}$ for the operator $O_{i}$ completely
destroys the outcome of measuring $\sigma _{\beta }^{k}\neq \sigma _{\alpha
}^{k}$ for other operators $O_{j}$. In principle, one could simulate the
required many-body interactions in AAPT (for each of the $4^{n}+1$
measurements) by a quantum circuit comprising about $O(n^{2})$ single and
two-qubit quantum operations (with the assumption of realizability of
non-local two-body interactions, i.e., with having access to two-qubit
interactions between every pairs of the $2n$ qubit system) \cite{MRL06}. For
a simple proof, we note that the measurement of an operator in the form $%
\sigma _{z}^{1}\sigma _{z}^{2}\sigma _{z}^{3}...\sigma _{z}^{2n}$ requires $%
2n$ sequential CNOT operations. For measuring a more general operator of the
form $O_{i=\{\alpha _{1},\alpha _{2},\alpha _{3,.}..\}}=\sigma _{\alpha
_{1}}^{1}\sigma _{\alpha _{2}}^{2}\sigma _{\alpha _{3}}^{3}...\sigma
_{\alpha _{2n}}^{2n}$, we need an additional $O(n)$ local single qubit
operations to make an appropriate change of basis. Therefore, for measuring $%
2n$ operators $O_{i}$, one at least needs to realize $(2n)^{2}$ quantum
operations. However, if only local two-body interactions are available
(i.e., if we are restricted to using nearest neighbor interactions) then $%
O(n^{3})$ single- and two-body quantum operations would be required. The
reason is that the overall number of operations grows by a factor of $O(n)$
due to the cost of transporting each two-qubit gate. A modified table for
comparing physical resources in different QPT methods is presented in Fig. 2.

\begin{figure}[tp]
\includegraphics[height=7.5cm,width=16.5cm]{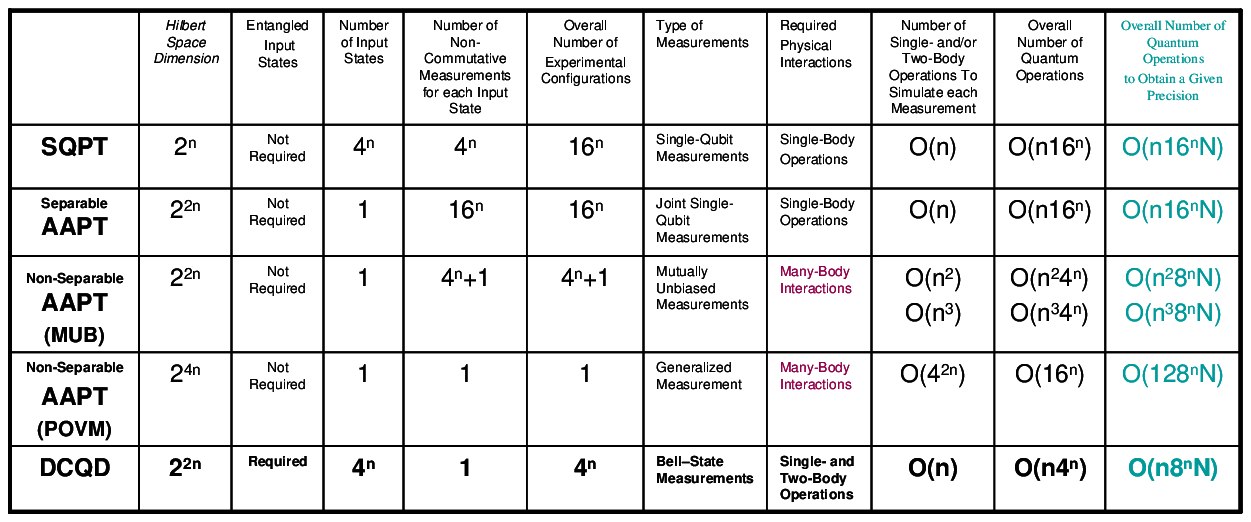}
\caption{Comparison of the required physical resources for characterizing an
arbitrary non-trace preserving CP quantum dynamical map on $n$ qubits. The
overall number of experimental configurations (5th column) is obtained from
multiplying the\ number of required input states and the number of
non-commuting measurements at the output states. In principle, the required
many-body interactions for AAPT+MUB can be simulated by a quantum circuit
comprising about $O(n^{2})$ [$O(n^{3})$] single and two-qubit quantum
operations with the assumption of non-local [local] two-body interactions.
However, the many-body interactions for a POVM in this context cannot be
efficiently simulated. The overall number of quantum operations (9th column)
is defined as the product of the number of experimental configurations and
the number of single- and/or two-body operations required for simulating
each measurement. Finally, the overall number of quantum operations for a
desired precision is obtained by multiplying by a factor of $2^{(k-1)n}N$
the 9th column, where $2^{kn}$ is the number of possible outcomes in each
QPT strategy, and $N$ is the number of repeated measurements to obtain the
precision $\protect\epsilon \sim \frac{1}{\protect\sqrt{N/2^{kn}}}$. We note
that DCQD is more efficient than all known QPT schemes with respect to the
overall number of quantum operations (see 9th and 10th columns).}
\end{figure}

Note that the AAPT plus MUB measurements with $O(n^{2})$ [or $O(n^{3})$]
two-body interactions is essentially in the same complexity class as a
quantum Fourier transform. This should be compared to the DCQD scheme with a
\emph{single step} of $n$ \textsc{CNOT} (between each qubit and its ancilla)
and $n$ Hadamard operations to realize a single Bell-state measurement. In
the context of estimating quantum dynamics, the implementation of $O(n^{2})$
or $O(n^{3})$ gates in AAPT is inefficient, since these operations not only
increase the time execution of each measurement, they also create additional
errors that would be very difficult to discriminate from the actual quantum
dynamical map. Moreover, the overall number of repetitions for these
operations scales poorly with a given desired precision.

Although MUB is not the most general measurement for state tomography of a $%
2n$-qubit system, it is well-understood (for systems whose dimensions is a
power of prime) to be the \emph{optimal} measurement scheme for such a task.
Therefore, any other measurement strategy\emph{\ within the same Hilbert
space} results in more ensemble measurements than MUB. For other systems
(whose dimensions is not a power of prime) the scaling of the AAPT
measurements becomes even worse; since in those systems, the existence and
construction of MUB is not fully understood and therefore in general one has
to measure a complete operator basis of $2n$-qubit system which has $%
2^{4n}-1 $ members.

In principle, one could devise intermediate strategies for AAPT, using
different combinations of single-, two-, and many-body measurements. The
number of measurements in such methods ranges from $4^{n}+1$ to $16^{n}$,
which is always larger than what is required in DCQD \cite{MohseniLidar06},
using $4^{n}$ Bell-state measurements. Therefore, in the $2^{2n}$
dimensional Hilbert space of the $2n$ system and ancillary qubits, DCQD
\emph{requires fewer experimental configurations than all other QPT schemes}.

Using DCQD one can in principle transfer $\log _{2}2^{2n}$ bits of classical
information between two parties, Alice and Bob, which is optimal according
to the Holevo bound \cite{HolevoBound}. Alice can realize this task by
encoding a string of $2^{2n}$ bits of classical information into a(n)
(engineered) quantum dynamics (e.g., by applying one of $2^{2n}$ unitary
operator basis to the $n$ qubits in her possession and then send them to
Bob). Bob can decode the message by a single measurement on $2n$ qubits
using DQQD scheme \cite{MohseniLidar06}. I.e., the overall number of
possible independent outcomes in each measurement in DCQD is $2^{2n}$, which
is exactly equal to the number of independent degrees of freedoms for a $2n$
qubit system, therefore,\emph{\ a maximum amount of information can be
extracted in each measurement in DCQD, which cannot be improved by any other
possible QPT\ strategies in the same Hilbert space}.

\subsection{AAPT with Generalized Measurements}

In principle, it is possible to perform the required quantum state
tomography at the output states of an AAPT scheme by utilizing a \textit{%
single} POVM or generalized measurement \cite{DArianoUQO02}. However, for
characterizing the dynamics on $n$ qubits the number of required ancillary
qubits should be increased from $n$ to $3n$ (see Fig. 3. of Ref \cite%
{DArianoUQO02}). This can be easily understood according to the Holevo
bound. For extracting complete information about a quantum dynamical map
(encoded by $2^{4n}$ independent parameters of the superoperator) in a
single measurement, one needs a Hilbert space of dimension at least $2^{4n}$%
; otherwise the Holevo bound cannot be satisfied. There are two major
disadvantages of using such a POVM compared to all other QPT schemes. (1)
The POVM measurement requires a general \emph{many-body interaction} between
$2n$ qubits that cannot be efficiently simulated. I.e., it requires an
exponential number of single- and two-qubits quantum operations. (2) The
number of required repetition of each measurement to obtain a desired
precision grows exponentially with $n$.

According to the general setting in Ref.~\cite{DArianoUQO02}, in
order to implement a POVM for extracting all the information about
any observable of an $n$-qubit system, one needs to realize a
global normal operator $\mathsf{H}$ (a \textit{single} universal
quantum observable) in the Hilbert-Schmidt space of the principle
system and an ancilla system in the form of
$\mathsf{H}=\sum_{i=1}^{2^{2n}}E_{i}\otimes P_{i}$. Here
$[\mathsf{H}, \mathsf{H}^{\dagger }]=0$, $\left\{ E_{i}\right\}$
is an operator basis for the $n$-qubit Hilbert space of the
system, and $\left\{P_{i}\right\}$ is a set of projections over
the ancilla Hilbert space (e.g., $P_i=|i\rangle \langle i|$, where
$\{|i\rangle\}$ is an orthonormal basis in the ancillary Hilbert
space). The operator $\mathsf{H}$ has the most general form of an
operator-Schmidt decomposition \cite{Nielsen} and cannot be
simulated in a polynomial number of steps. It is known that in
general at least $O(4^{2n})$ single- and two-qubit operations are
needed to simulate such general many-body operations acting on
$2n$ qubits \cite{Vivek04} (see also \cite{Nielsen} for different
measures of complexity of a given quantum dynamics).

One important disadvantage of all QPT schemes in a larger Hilbert space
(that rely on calculating $2^{kn}$ different joint probability
distributions) is that each measurement has to be repeated by a factor $%
2^{(k-1)n}$ in order to build the same statistics as a SQPT scheme. Let us
consider an implementation of SQPT with a desired precision of $\epsilon
\sim \frac{1}{\sqrt{N}}$ in characterizing parameters of a superoperator,
where $\epsilon $ represents the standard deviation and $N$ \ is the number
of repeated measurements. Since each measurement in SQPT has $2^{n}$
possible outcomes, the precision $\epsilon $ can be obtained by $N=\frac{%
2^{n}}{\epsilon ^{2}}$ measurements. Note that in order to obtain a similar
statistical error $\epsilon $ with other methods of QPT, with $2^{kn}$
possible outcomes, we need to perform $N^{\prime }=2^{(k-1)n}N$ measurements
for each experimental configuration. Therefore, the actual number of
measurements for a POVM strategy, with $2^{4n}$ possible outcomes, grows by
a factor of $2^{3n}$\ with respect to SQPT and $2^{2n}$ with respect to DCQD
-- see the last column of the table in Fig. 2. We note that the overall
number of quantum operations is still optimal for DCQD for any desired
precision.

\section{Reply to the second comment: Independence from AAPT and usage of
quantumness}

Here, we argue that the DCQD scheme is an independent algorithm from the
AAPT scheme. First, we note that DCQD has a different scaling from the AAPT
scheme in the sense of the overall number of experimental configurations, or
similar scaling while using only two-body interactions compared to many-body
interactions in the non-separable AAPT. Clearly, such results could not have
been obtained if DCQD were merely a subclass of AAPT.

Second, the required input states for DCQD \emph{must be entangled}, which
is in complete contrast to AAPT, where entanglement is not required at the
input level. E.g., for characterizing quantum dynamical populations \cite%
{MohseniLidar06}, the input state in DCQD must be maximally entangled in
order to form a nondegenerate stabilizer state (with two independent
stabilizer generators). This follows from the quantum Hamming bound,
according to which only nondegenerate stabilizer states can be utilized for
obtaining full information about the nature of all $2^{2}$ different error
operator basis elements, acting on a single qubit of a two-qubit system \cite%
{Nielsen:book}. For characterizing quantum dynamical coherence, the
entanglement in input states is absolutely necessary, for otherwise the
expectation values of the normalizers always vanish and therefore do not
provide any information about the dynamics. In addition, the error-detection
measurements in DCQD (e.g., $Z^{A}Z^{B}$, $X^{A}Z^{B}$) are fundamentally
non-separable, which is again in contrast to AAPT which also can be
performed by joint single-qubit measurements.

Third, the DCQD method utilizes a different methodological approach to
quantum dynamical characterization than the AAPT schemes. DCQD utilizes
\emph{a set of entangled \ }input states and provides the set of \emph{%
commuting} observables to \emph{maximize} the amount of classical
information about the dynamics, $\log _{2}d^{2},$ that can be obtained at
each output state (according to the Holevo bound), without completely
characterizing any of the output states. However, AAPT utilizes \emph{a
single faithful} (not necessary entangled) input state and provides the
\emph{minimal} set of \emph{non-commuting }observables that should be
measured in order to \emph{completely characterize the output state} of the
combined system and ancilla, and the amount of classical information that
can be obtain in each measurement, $\log _{2}(d^{2}-1)$, is less than the
maximum allowable according to the Holevo bound ($\log _{2}d^{2}$).
Therefore, there cannot be any direct correspondence between DCQD and AAPT
in any fixed Hilbert space of the system and ancilla. We believe that the
only true similarity between AAPT and DCQD is the fact that both methods
utilize the degrees of freedoms of an auxiliary system.

\section{Reply to the third comment: Direct characterization of dynamics}

In order to remove any ambiguity, we would like to define what we mean by
\emph{direct} characterization of quantum dynamics. We define a QPT method
to be a direct method if it satisfies these two conditions: (1) It should
not rely on complete state tomography of the output states. (2) Each
experimental outcome (joint probability distribution of observables) give
direct information about either a single element of the superoperator, ($%
\chi _{mm}$)\ or a specific known subset of the superoperator's elements
(e.g., $\chi _{mm},\chi_{nn},\text{Re}(\chi _{mn})$).

AAPT is not a direct method because it does not satisfy the condition (1).
One could argue that when the local state of the ancilla is known ($d^{2}$
parameters), only $d^{4}-d^{2}$ additional parameters must be characterized.
However, since in this case one eventually also has access to $d^{4}$
parameters ($d^{2}$ of which were known from the beginning, the rest
measured), this should clearly also count as complete state tomography

\section{Conclusion}

We agree that demonstrating an \emph{absolute} advantage of one QPT scheme
over the other QPT methods requires a complete quantification of the
complexity of the preparations and measurements procedures. Indeed, we
present a more detailed analysis in a separate publication \cite{MRL06}. In
conclusion, we believe the following statements are true:

\begin{enumerate}
\item DCQD is a quantum algorithm for complete and direct characterization
of quantum dynamics, which does not require state tomography.

\item We have proved that DCQD is optimal in the sense of both the \emph{%
required number of experimental configurations} and the \emph{number of
possible outcomes}, over all other known QPT schemes in a given Hilbert
space.

\item DCQD is \textit{quadratically} more efficient than all separable QPT
schemes in the number of \emph{experimental configurations}.

\item A similar scale-up in the number of experimental configurations is
achievable with the AAPT scheme and MUB measurements only if \emph{many-body
interactions} are realized (or simulated with $O(n^{2})$ or $O(n^{3})$
single- and two-body gates).

\item In principle, by utilizing POVMs, a \textit{single} experimental
configuration is sufficient for a complete QPT, however, one should realize
many-body interactions that are not experimentally available and cannot be
efficiently simulated by single- and two-body interactions. Moreover, the
POVM strategy has to be repeated as many as $4^{n}$ times more than DCQD to
obtain a similar precision.

\item DCQD is new method for QPT, and cannot be considered merely as a
subclass of any known QPT methods.

\item DCQD is the first theory that utilizes quantum error detection methods
in quantum process tomography.
\end{enumerate}

\section{Applications and future work}

We believe that a potentially important advantage of DCQD is for use in
partial characterization of quantum dynamics, where we cannot afford or do
not need to carry out a full characterization of the quantum system under
study, or when we have some a priori knowledge about the dynamics. We have
already presented two examples in connection with simultaneous measurement
of $T_{1}$ and $T_{2}$, and realization of generalized quantum dense coding
tasks. Other implications and applications of DCQD remain to be investigated
and explored, specifically, for obtaining a polynomial scale-up in physical
resources for partial characterization of quantum dynamics. We believe that
in some specific regimes DCQD could have near-term applications (within the
next 5-10 years) for complete verification of small quantum information
processing units (fewer than five qubits or so), especially in trapped-ion
and liquid-state NMR systems. For example, the number of required
experimental configurations for systems of 3 or 4 physical qubits is reduced
from $5000$ and $65000$ (in SQPT) to $64$ and $256$, respectively, in our
scheme. Complete characterization of such dynamics would be essential for
verification of quantum key distribution procedures, teleportation units (in
both quantum communication and scalable quantum computation), quantum
repeaters, quantum error correction procedures, and more generally, in any
situation in quantum physics where a few qudits have a common local bath and
interact with each other. Another interesting extension of DCQD is to
develop a theory for closed-loop and continuous characterization of quantum
dynamics by utilizing weak measurements for our error-detection procedures.

This work was supported by the Natural Sciences and Engineering Research
Council of Canada (to M.M. and D.A.L.), DARPA-QuIST, and the Sloan
Foundation (to D.A.L). We thank M. Ziman for inspiring us to more carefully
consider the resource requirements in our DCQD\ scheme and other known QPT\
schemes, and also acknowledge many useful discussions with A. T. Rezakhani.


\begin{thebibliography}{99}
\bibitem{MarioNote06} M. Ziman, Notes on optimality of direct
characterization of quantum dynamics, quant-ph/0603151.

\bibitem{MohseniLidar06} M. Mohseni and D. A. Lidar, Direct characterization
of quantum dynamics: I. General theory, quant-ph/0601033, Direct
characterization of quantum dynamics: II. Detailed analysis,
quant-ph/0601034.

\bibitem{Nielsen:book} {M. A. Nielsen and I. L. Chuang}, \textit{Quantum
Computation and Quantum Information} (Cambridge University Press, Cambridge,
UK, 2001).

\bibitem{DAriano01} G. M. D'Ariano and P. Lo Presti, Phys. Rev. Lett.
\textbf{86}, 4195 (2001).

\bibitem{Wooters:89} W. K. Wootters and B. D. Fields, Ann. Phys. \textbf{191}%
, 363 (1989).

\bibitem{LawrenceMUB02} J. Lawrence, C. Brukner, Phys, Rev. A. \textbf{65},
032320 (2002).

\bibitem{DArianoUQO02} G. M. D'Ariano, Phys. Lett. A \textbf{300}, 1 (2002).

\bibitem{MRL06} M. Mohseni, A. T. Rezakhani, and D. A. Lidar, in preparation
(2006).

\bibitem{HolevoBound} A. S. Holevo, Probl. Infor. Transm. \textbf{9}, 110
(1973).

\bibitem{Nielsen} M. A. Nielsen \textit{et al.}, Phys. Rev. A \textbf{67},
052301 (2003).

\bibitem{Vivek04} V. V. Shende \textit{et al.}, Phys. Rev. A \textbf{69},
062321 (2004); M. M\"{o}tt\"{o}nen \textit{\ et al.}, Phys. Rev. Lett.
\textbf{93}, 130502 (2004).
\end{thebibliography}
\end{document}